\documentclass[11pt,a4paper]{article}
\usepackage{amsfonts}
\usepackage{amssymb}
\usepackage[centertags]{amsmath}
\usepackage{graphicx}
\usepackage{booktabs}
\usepackage{theorem}
\usepackage{array}
\usepackage{colordvi}
\usepackage{ulem}
\usepackage[small,bf]{caption}
\usepackage{cite}
\usepackage{color}

\allowdisplaybreaks[1]

\numberwithin{equation}{section}

\newcommand{\be}{\begin{equation}}
\newcommand{\ee}{\end{equation}}
\newcommand{\beq}{\begin{equation}}
\newcommand{\eeq}{\end{equation}}
\newcommand{\bea}{\begin{eqnarray}}
\newcommand{\eea}{\end{eqnarray}}

\newcommand{\lv}{\langle}
\newcommand{\rv}{\rangle}

\makeatletter

\newcommand{\Rmnum}[1]{\expandafter\@slowromancap\romannumeral #1@}
\makeatother

\begin{document}

\begin{titlepage}

\vspace*{-15mm}
\begin{flushright}
MPP-2011-101\\
SHEP-11-23\\
SISSA 44/2011/EP
\end{flushright}
\vspace*{0.7cm}

\begin{center}
{
\bf\LARGE
Trimaximal mixing with predicted $\boldsymbol{\theta_{13}}$
from a new type of constrained sequential dominance}
\\[8mm]
Stefan~Antusch$^{\star}$
\footnote{E-mail: \texttt{stefan.antusch@unibas.ch}},
Stephen~F.~King$^{\dagger}$
\footnote{E-mail: \texttt{king@soton.ac.uk}},
Christoph~Luhn$^{\dagger}$
\footnote{E-mail: \texttt{christoph.luhn@soton.ac.uk}},
Martin~Spinrath$^{\ddag}$
\footnote{E-mail: \texttt{spinrath@sissa.it}},
\\[1mm]
\end{center}
\vspace*{0.50cm}
\centerline{$^{\star}$ \it
 Department of Physics, University of Basel,}
\centerline{\it
Klingelbergstr.~82, CH-4056 Basel, Switzerland}
\vspace*{0.2cm}
\centerline{$^{\star}$ \it
Max-Planck-Institut f\"ur Physik (Werner-Heisenberg-Institut),}
\centerline{\it
F\"ohringer Ring 6, D-80805 M\"unchen, Germany}
\vspace*{0.2cm}
\centerline{$^{\dagger}$ \it
School of Physics and Astronomy, University of Southampton,}
\centerline{\it
SO17 1BJ Southampton, United Kingdom }
\vspace*{0.2cm}
\centerline{$^{\ddag}$ \it
SISSA/ISAS and INFN,}
\centerline{\it
Via Bonomea 265, I-34136 Trieste, Italy }
\vspace*{1.20cm}
\begin{abstract}

\noindent
Following the recent T2K indication of a sizeable reactor angle, we present a class of models
which fix $\theta_{13}$ while preserving trimaximal solar mixing. The
models are based on a new type of constrained sequential dominance involving new
vacuum alignments, along the $(1,2,0)^T$ or $(1,0,2)^T$ directions in flavour
space. We show that such alignments are easily achieved using orthogonality,
and may  replace the role of the subdominant flavon alignment $(1,1,1)^T$ in
constrained sequential dominance. In such models, with a normal hierarchical
spectrum, the reactor angle is related to a ratio of neutrino masses by
$\theta_{13} = \frac{\sqrt{2}}{3}\frac{m^\nu_2}{m^\nu_3}$, leading to
$\theta_{13} \sim 5^\circ - 6^\circ$, while the atmospheric angle is given by
the sum rule $\theta_{23} \approx 45^\circ + \sqrt{2} \theta_{13} \cos
\delta$.
We find that leptogenesis is unsuppressed due to the violation of form dominance and 
that the CP violating phase responsible for leptogenesis is precisely equal to the Dirac CP phase $\delta$,
providing a direct link between leptogenesis and neutrino mixing in this class of models.
\end{abstract}

\end{titlepage}

\setcounter{footnote}{0}

\section{Introduction}
Recently T2K have published evidence for a large non-zero reactor angle
\cite{Abe:2011sj} which, when combined with data from MINOS and other
experiments in a global fit, yields \cite{Fogli:2011qn,Schwetz:2011zk}
\be
6^\circ \lesssim \theta_{13} \lesssim  9^\circ ,
\label{reactor}
\ee
with a statistical significance of a non-zero reactor angle of about 3$\sigma$.
If confirmed this would rule out the hypothesis of exact tri-bimaximal (TB)
mixing~\cite{Harrison:2002er}, and a flurry of alternative proposals have
recently been put forward \cite{flurry}.

For example, an attractive scheme based on trimaximal (TM) mixing remains viable 
\cite{Haba:2006dz}. TM mixing is defined to maintain the second column of the
TB mixing matrix and hence preserves the solar mixing angle prediction
$\sin\theta_{12} \approx 1/\sqrt{3}$. However there is another variation of TM
mixing which also preserves this good solar mixing angle prediction by
maintaining the first column of the TB matrix, namely TM$_1$ mixing
\cite{Lam:2006wm}: 
\begin{equation}
\label{TM1}
U_{\mathrm{TM_1}}~=~P'\begin{pmatrix} 
\frac{2}{\sqrt{6}}
&\frac{1}{\sqrt{3}}\cos\vartheta 
&\frac{1}{\sqrt{3}}\sin\vartheta \,e^{i\rho}\\
-\frac{1}{\sqrt{6}}
&\frac{1}{\sqrt{3}}\cos\vartheta -\frac{1}{\sqrt{2}}\sin\vartheta\,e^{-i\rho}
& \phantom{-}\frac{1}{\sqrt{2}}\cos\vartheta+\frac{1}{\sqrt{3}}\sin\vartheta  \,e^{i\rho} \\ 
-\frac{1}{\sqrt{6}}
& \frac{1}{\sqrt{3}}\cos\vartheta+ \frac{1}{\sqrt{2}}\sin\vartheta\,e^{-i\rho}
& - \frac{1}{\sqrt{2}}\cos\vartheta+\frac{1}{\sqrt{3}}\sin\vartheta \,e^{i\rho} 
\end{pmatrix} P \ ,
\end{equation}
where $\frac{1}{\sqrt{3}} \sin \vartheta=\sin\theta_{13}$, $P'$ is a diagonal
phase matrix required to put $U_{\mathrm{TM_1}}$ into the PDG convention
\cite{Nakamura:2010zzi}, and $P={\rm diag}(1,e^{i\frac{\alpha_{2}}{2} },
e^{i\frac{\alpha_{3}}{2} })$ contains the usual Majorana phases. In particular
TM$_1$ mixing approximately preserves the successful TB mixing for the solar
mixing angle $\theta_{12}\approx 35^\circ$ as the correction due to a non-zero
but relatively small reactor angle is of second order. Although TM$_1$ mixing
reduces to TB mixing in the limit that $\vartheta \rightarrow 0$, it is worth
emphasising that in general TM$_1$ mixing involves an undetermined reactor
angle $\theta_{13}$ which could in principle be large or even maximal
(e.g. $45^\circ$). The observed smallness of the reactor angle~$\theta_{13}$
compared to the atmospheric angle $\theta_{23}\approx 45^\circ$ and the solar
angle $\theta_{12}\approx 34^\circ$  \cite{Fogli:2011qn} is therefore not
explained by the TM$_1$ hypothesis alone. Clearly the relative smallness of
the reactor angle can only be explained with additional model dependent input. 
Although there are models of TM mixing which can account for the smallness 
of the reactor angle \cite{King:2011zj} so far there is no model in the literature
for TM$_1$ mixing, let alone one which fixes the reactor angle.

In this paper we propose a model of TM$_1$ mixing where the magnitude of the
reactor angle is fixed. The model we discuss is actually representative of a
general strategy for obtaining TM$_1$ mixing using sequential dominance (SD)
\cite{King:1998jw} and vacuum alignment. The strategy of combining SD with
vacuum alignment is familiar from the constrained sequential dominance (CSD)
approach to TB mixing \cite{King:2005bj} where a neutrino mass hierarchy is
assumed and the dominant and subdominant flavons responsible for the
atmospheric and solar neutrino masses are aligned in the directions of the 
third and second columns of the TB mixing matrix, namely 
$\lv\phi_1^{\nu}\rv \propto  (0,1,-1)^T$ and 
$\lv\phi_2^{\nu}\rv\propto (1,1,1)^T$.
The new idea here is to maintain the usual vacuum alignment for the dominant
flavon,  $\lv\phi_1^{\nu}\rv\propto (0,1,-1)^T$ as in CSD, but to replace the
effect of the subdominant flavon vacuum alignment by a different one, namely
either $\lv\phi_{120}\rv\propto (1,2,0)^T$ or $\lv\phi_{102}\rv\propto (1,0,2)^T$,
where such alignments may be naturally achieved from the standard ones using
orthogonality arguments. We shall refer to this new approach as CSD2. We shall
show that CSD2 leads to TM$_1$ mixing and a reactor angle which, at leading
order, is predicted to be proportional to the ratio of the solar to the
atmospheric neutrino masses, $\theta_{13} = \frac{\sqrt{2}}{3} \,
\frac{m^\nu_2}{m^\nu_3}$.   

It is interesting to compare the predictions of CSD2 to another alternative to CSD that 
has been proposed to account for a reactor angle called partially constrained
sequential dominance (PCSD) \cite{King:2009qt}. PCSD involves a vacuum
misalignment of the dominant flavon alignment to $(\varepsilon ,1,-1)^T$, with a 
subdominant flavon alignment $(1,1,1)^T$, leading to tri-bimaximal-reactor
(TBR) mixing \cite{King:2009qt} in which only the reactor angle is switched
on, while the atmospheric and solar angles retain their TB values. However, in
the case of PCSD, the value of the reactor angle is not predicted whereas CSD2 
leads to the above relation.

The layout of the rest of the paper is as follows. In Section~\ref{2} we
describe CSD2 including a discussion of vacuum alignment and an example of a
model based on CSD2. In Section~\ref{3} we discuss the phenomenology of CSD2,
first showing that it reproduces TM$_1$ mixing exactly,
then comparing the second order analytic results to a numerical treatment. In
Section~\ref{4} we show that leptogenesis is unsuppressed and moreover CSD2
leads to a link between the CP phase for leptogenesis and the Dirac CP
phase~$\delta$. We conclude in Section~\ref{5}.

\section{\label{2}A new type of constrained sequential dominance}

\subsection{CSD and TB mixing}

Assuming the type I see-saw mechanism, in the diagonal right-handed (RH)
neutrino mass basis we may write $M_R={\rm diag}(M_A, M_B, M_C)$ and the neutrino
Yukawa matrix as $Y_{\nu} = (A, B, C)$ where $A, B, C$  are three column
vectors. Then the type I see-saw formula
${M_{\nu}}=Y_{\nu}M_{R}^{-1}Y_{\nu}^T$ gives
\begin{equation}
\label{mLLCSD} {M_{\nu}}= \frac{v^2 A A^T}{M_A}+ \frac{v^2 B B^T}{M_B} +
\frac{v^2 C C^T}{M_C}\ . 
\end{equation}
SD corresponds to a hierarchy of contributions 
$ \frac{A A^T}{M_A}\gg \frac{B B^T}{M_B} \gg \frac{C C^T}{M_C}$ 
corresponding to the physical neutrino mass hierarchy  
$m^\nu_3 \gg m^\nu_2 \gg m^\nu_1$ \cite{King:1998jw}.
The Yukawa couplings in $A$ determine the atmospheric angle, while those in
$B$  determine the solar angle, with the reactor angle dependent on both
\cite{King:1998jw}. For a strong hierarchy one can ignore the Yukawa couplings
in $C$ and the effect of the third right-handed  neutrino. 

TB mixing naturally emerges from CSD \cite{King:2005bj} where the neutrino
Yukawa matrix $Y_\nu$ and the right-handed neutrino mass matrix $M_R$ take the
constrained form 
\begin{equation}
Y_\nu \,=\, \begin{pmatrix} 0 & b \\ a &b \\ -a & b \end{pmatrix}, \qquad M_R
\,=\, \begin{pmatrix} M_{A} & 0 \\ 0 & M_{B}  \end{pmatrix}, 
\end{equation}
assuming $m^\nu_1=0$. In models with non-Abelian family symmetries this
structure can be explained by two flavons pointing in the directions in
flavour space defined by the columns of $Y_\nu$:
\begin{equation}
\lv\phi^\nu_{1}\rv \propto \begin{pmatrix} 0 \\ 1 \\ -1 \end{pmatrix} , \qquad 
\lv\phi^\nu_{2} \rv \propto \begin{pmatrix} 1 \\ 1 \\ 1 \end{pmatrix}.
\end{equation}

\subsection{CSD2 and vacuum alignment}

As our starting point, we consider the vacuum alignment sector of any SD
flavour model where the flavons, which furnish triplet representations under a
family symmetry $G_F$, are typically aligned in the three orthogonal
directions 
\begin{equation} \label{eq:nuDirections}
\lv\phi^{\nu}_{1}\rv \propto \begin{pmatrix} 0 \\ 1 \\ -1 \end{pmatrix} , \quad 
\lv\phi^{\nu}_{2}\rv \propto \begin{pmatrix} 1 \\ 1 \\ 1 \end{pmatrix}  , \quad 
\lv\phi^{\nu}_{3}\rv \propto \begin{pmatrix} -2 \\ 1 \\ 1 \end{pmatrix}  .
\end{equation}
Applying these flavons to build up  the neutrino Yukawa matrix, one would
obtain TB neutrino mixing and a form diagonalisable neutrino mass matrix.

In many models three additional orthogonal flavons are present,
\begin{equation} \label{eq:eDirections}
\lv\phi^e_{1}\rv \propto \begin{pmatrix} 1 \\ 0 \\  0 \end{pmatrix} , \quad 
\lv\phi^e_{2}\rv \propto \begin{pmatrix} 0 \\ 1 \\ 0 \end{pmatrix}  , \quad  
\lv\phi^e_{3}\rv \propto \begin{pmatrix} 0 \\ 0 \\ 1 \end{pmatrix} .  
\end{equation}
The third flavon $\phi^e_3$ is generically introduced to generate hierarchical
quark and charged lepton mass matrices, and typically governs the masses of
the third generation of charged particles.

In this paper we augment this set of CSD flavons by another flavon pointing in
the direction 
\begin{equation}
\lv\phi_{120} \rv\propto \begin{pmatrix} 1 \\ 2 \\ 0 \end{pmatrix} 
\quad  \text{or} \quad 
\lv\phi_{102} \rv\propto  \begin{pmatrix} 1 \\ 0 \\ 2 \end{pmatrix} .
\end{equation}
These alignments can be obtained by requiring the orthogonality conditions
$\lv\phi_{120}\rv \cdot\lv \phi^{\nu}_{3}\rv = \lv\phi_{120}\rv \cdot\lv \phi^e_{3}\rv =0$ 
or alternatively 
$\lv\phi_{102}\rv\cdot\lv \phi^{\nu}_{3}\rv=\lv\phi_{102}\rv\cdot\lv\phi^e_{2} \rv=0$, 
where the ``$\cdot$''  denotes the usual $SO(3)$ inner product. Implicitly we
assume real triplet representations, but it is straightforward to extend  this
mechanism to complex representations. 
The orthogonality conditions can be realised easily with ``Lagrange
multiplier'' superfields $D_1$ and $D_2$, which are singlets under the family
symmetry $G_F$. They give the following terms in the superpotential  
\begin{equation} \label{eq:NewAlignment}
D_1 (\phi_{120} \cdot \phi^{\nu}_{3}) + D_2 (\phi_{120} \cdot \phi^e_{3}) \quad \text{or} \quad D_1 (\phi_{102} \cdot \phi^{\nu}_{3}) + D_2 (\phi_{102} \cdot \phi^e_{2}) \;.
\end{equation}
The $F$-term conditions $|F_{D_1}| = |F_{D_2}| = 0$ are equivalent to the
orthogonality conditions and therefore yield the desired alignments. 

The shaping symmetries of a specific model select the flavons which couple to
the RH neutrinos and thus determine the structure of the neutrino Yukawa
matrix. CSD2 corresponds to the case where the dominant flavon is $\phi^\nu_1$ as in
CSD, and the subdominant flavon is taken to be either $\phi_{120}$ or $\phi_{102}$.
In the following subsection, we sketch a model in which this is achieved.

\subsection{An example of a model with CSD2}

In the following we briefly outline how to implement the discussed alignment
into a concrete model of lepton masses and mixing angles.
As the model is mainly for the purpose of illustration, we do not discuss the quark sector and the charged lepton mass matrix is diagonal by construction.

\begin{table}
\centering
\begin{tabular}{lccccccccc} \toprule
  & $SU(2) \times U(1)$& $A_4$ & $Z_4^{(1)}$ & $Z_4^{(2)}$ & $Z_4^{(3)}$ & $Z_4^{(4)}$ & $Z_4^{(5)}$ & $Z_4^{(6)}$ & $Z_4^{(7)}$ \\
\midrule
$L$ & $\mathbf{2}_{-1}$ & $\mathbf{3}$ \\
$e^c$ & $\mathbf{1}_{2}$ & $\mathbf{1}$ & 3 \\
$\mu^c$ & $\mathbf{1}_{2}$ & $\mathbf{1}$ & & 3 \\
$\tau^c$ & $\mathbf{1}_{2}$ & $\mathbf{1}$ & & & 3 \\
$N_1$ & $\mathbf{1}_{0}$ & $\mathbf{1}$ & & & & 3  \\
$N_2$ & $\mathbf{1}_{0}$ & $\mathbf{1}$ & & & & & 3 \\
$H_1$ & $\mathbf{2}_{-1}$ & $\mathbf{1}$ \\
$H_2$ & $\mathbf{2}_{1}$ & $\mathbf{1}$ \\
\midrule
$\phi^e_1$ & $\mathbf{1}_{0}$ & $\mathbf{3}$ & 1 \\
$\phi^e_2$ & $\mathbf{1}_{0}$ & $\mathbf{3}$ & & 1 \\
$\phi^e_3$ & $\mathbf{1}_{0}$ & $\mathbf{3}$ & & & 1 \\
$\phi^{\nu}_1$ & $\mathbf{1}_{0}$ & $\mathbf{3}$ & & & & 1 & \\
$\phi^{\nu}_2$ & $\mathbf{1}_{0}$ & $\mathbf{3}$ & & & & & & 1  \\
$\xi$ & $\mathbf{1}_{0}$ & $\mathbf{1}$ & & & & & & 1 \\
$\phi^{\nu}_3$ & $\mathbf{1}_{0}$ & $\mathbf{3}$ & & & & & & & 1 \\
$\phi_{120}$ & $\mathbf{1}_{0}$ & $\mathbf{3}$ & & & & & 1 \\
$\phi_{102}$ & $\mathbf{1}_{0}$ & $\mathbf{3}$ & & & & & 1 \\
\bottomrule
\end{tabular}
\caption{The symmetries of the example model of lepton masses and
  mixings. Only the matter, the Higgs and the flavon fields are shown. The
  charges of the driving fields can be easily inferred from the corresponding
  superpotential terms.\label{Tab:ToyModel}}
\end{table}

We start with a short discussion on how to achieve the alignments in
Eqs.~\eqref{eq:nuDirections} and \eqref{eq:eDirections}. This is rather
standard and we follow closely the alignment as discussed in
\cite{Antusch:2011sx}. Although we do not implement spontaneous CP violation
we stick to $Z_4$ symmetries to make the analogy more obvious. 

The vacuum alignment of $\phi^{e}_{1,2,3}$ is generated by the renormalisable
superpotential 
\begin{equation}
 \mathcal{W}_e \sim A^{e}_i (\phi^{e}_{i} \star \phi^{e}_{i}) + O^{e}_{ij} (\phi^{e}_{i} \cdot \phi^{e}_{j} ) \;,
\end{equation}
where ``$\star$'' denotes the symmetric cross product in $A_4$, which is defined
as $(x \star y)_i = s_{ijk} x^j y^k$, with $s_{ijk} = + 1$ on all permutations
$\{i,j,k\} \in \{1,2,3\}$ and zero otherwise, see also \cite{King:2006np}.
Here and in the rest of this section we drop coupling
constants in the superpotential to increase the readability. The $F$-term
conditions of the triplet driving fields $A_i^{e}$ align the flavon vevs in the desired
directions\footnote{At this stage the flavon vevs could also vanish. We assume
  that they are driven to non-zero values by either soft SUSY breaking mass
  terms or higher-dimensional terms in the superpotential.} and the singlet
driving fields $O^{e}_{ij}$ enforce the flavons to point in three different directions. 
Note that  $A_i^{e}$ is doubly charged under $Z_4^{(i)}$ while $O^{e}_{ij}$
carries a charge of three under both $Z_4^{(i)}$ and $Z_4^{(j)}$. We also
assume a $U(1)_R$ symmetry under which the matter fields carry a charge of
one, the flavons are neutral and the driving fields are doubly charged. 

For the alignment of  the $\phi^{\nu}$ flavons we employ the superpotential
\begin{equation}
 \mathcal{W}_\nu \sim 
A_2^{\nu} \left( \xi \phi_2^\nu + \phi_2^\nu \star \phi_2^\nu \right) + 
O^{\nu}_{ij} (\phi^{\nu}_{i} \cdot \phi^{\nu}_{j}) +
O^{\nu e}_{11} ( \phi_1^\nu \cdot \phi_1^e) \;,
\end{equation}
where we use a similar notation as before and the charges under the $A_4$ and
$Z_4$ symmetries are distributed in the same way. Note that we have introduced
an additional auxiliary singlet flavon field $\xi$ with a non-zero vev to align
$\phi_2^{\nu}$. Having obtained the alignment $\lv\phi_2^{\nu}\rv$ from the
$A_2^{\nu}$ driving field, the remaining neutrino flavons $\phi_1^\nu$ and
$\phi_3^\nu$ are aligned by the driving fields $O^{\nu e}_{11}$ and
$O^{\nu}_{ij}$.

Eventually, the new flavon alignments are achieved by adding the two singlet
driving fields $D_1$ and $D_2$ of Eq.~\eqref{eq:NewAlignment} as discussed
previously. $D_1$ is charged under $Z_4^{(5)}$ and $Z_4^{(7)}$ while $D_2$ is
charged under $Z_4^{(5)}$ and $Z_4^{(2/3)}$ for $\phi_{102/120}$. 

With the symmetries and the field content as given in Tab.\ \ref{Tab:ToyModel}
we end up with the following Yukawa superpotential
\begin{equation}
\mathcal{W}_{\text{Yuk}} \sim \frac{1}{\Lambda} \left( \phi^e_1 \cdot L e^c
H_1 +  \phi^e_2\cdot L \mu^c  H_1 +  \phi^e_3 \cdot L \tau^c  H_1 +
\phi^{\nu}_1 \cdot L N_1 H_2 + \phi_{120/102} \cdot L N_2 H_2 \right),
\end{equation}
which give the Yukawa couplings after the flavons develop their
vevs. $\Lambda$ is a generic messenger mass scale. The charged lepton Yukawa
matrix is diagonal due to the alignment of the $\phi^e$ flavons: 
\begin{equation}
Y_e = \text{diag}(y_e, y_\mu, y_\tau) \;,
\end{equation}
and the neutrino Yukawa couplings $Y_\nu = (A,B)$ read
\begin{equation} \label{eq:NeutrinoYukawas}
Y^{(120)}_\nu = \begin{pmatrix} 0 & b \\ a & 2 b \\ -a & 0 \end{pmatrix}
\quad \text{or} \quad Y^{(102)}_\nu =
 \begin{pmatrix} 0 & b \\ a & 0 \\ -a & 2 b \end{pmatrix} \;, 
\end{equation}
depending on the choice of the subdominant flavon, either $\phi_{120}$ or
$\phi_{102}$. The parameters $a$ and~$b$ can be determined from the parameters
in the superpotential.
Later on we will see, that a relative phase difference $\arg (a/b) = 45^\circ, 135^\circ, 225^\circ$ or $315^\circ$, which translates into a Dirac CP phase $\delta = 90^\circ$ or $270^\circ$, is preferred by experimental data and that this would also maximise the generated baryon asymmetry. Such a phase difference can be easily obtained in the context of spontaneous CP violation from discrete symmetries as discussed in \cite{Antusch:2011sx}, which could be applied here straightforwardly.

Due to the $Z_4$ symmetries the RH neutrinos have no mass terms at the
renormalisable level, but they become massive after the flavons develop their
vevs due to the following terms in the superpotential 
\begin{equation}
 \mathcal{W}_R \sim \frac{1}{\Lambda} ( \phi_1^\nu )^2 N_1^2 + \frac{1}{\Lambda} \phi_{120/102}^2 N_2^2 \;.
\end{equation} From the symmetries alone also terms like $\phi_1^\nu \cdot
\phi_{120/102} N_1 N_2$ would be allowed, but we assume, that the messenger fields
mediating such operators are absent. Under this common assumption the RH neutrino
mass matrix is diagonal 
\begin{equation} \label{eq:MR}
 M_R  =  \begin{pmatrix} M_{A} & 0 \\ 0 & M_{B} \end{pmatrix} \;.
\end{equation}

\section{\label{3}The phenomenology of CSD2}

\subsection{Predictive trimaximal mixing from CSD2}
With the charged lepton mass matrix being diagonal, the Pontecorvo--Maki--Nakagawa--Sakata (PMNS) mixing originates 
solely from the neutrino sector. As discussed in the previous section we
introduce two RH neutrinos $N_i$ ($i=1,2$) which entails one massless light
neutrino. The RH neutrino mass matrix $M_R$ is assumed to be diagonal and each
$N_i$ couples to its own flavon. Adopting $\phi^\nu_1$ for the dominant and
$\phi_{120}$ for the subdominant term, the resulting neutrino Yukawa
matrix is $Y_\nu^{(120)}$, see Eq.~\eqref{eq:NeutrinoYukawas}.\footnote{We
  comment on the case where the subdominant flavon is taken to be $\phi_{102}$
  below   Eqs.~(\ref{analytic1}-\ref{analytic2}).
  Note that this resembles the case with two-texture zeros in the neutrino Yukawa matrix,
  whose phenomenology was extensively discussed in \cite{Ibarra:2003xp, Ibarra:2003up}.
  Here we go beyond those papers by giving an explicit vacuum alignment mechanism for
  such a texture, where all mixing angles and phases depend on a single complex parameter.
  }
Then the (type-I) seesaw
formula leads to a simple effective light neutrino mass matrix, given by
\begin{equation}
\begin{split}
M_\nu &= \frac{v^2 {A A^T}}{M_A} 
+ \frac{v^2 {B B^T}}{M_B}
= m_a \begin{pmatrix} 0 & 0 & 0 \\ 0 & 1 & -1 \\ 0 & -1 & 1 \end{pmatrix} + m_b \begin{pmatrix} 1 & 2 & 0 \\ 2 & 4 & 0 \\ 0 & 0 & 0  \end{pmatrix} \\
&= 
 m_a \begin{pmatrix} 
\frac{m_b}{m_a}&2 \frac{m_b}{m_a} & 0  \\ 
2 \frac{m_b}{m_a}& 1+ 4 \frac{m_b}{m_a}  & - 1 \\ 
0 & - 1 & 1 
\end{pmatrix}, \label{numassmatrix}
\end{split}
\end{equation}
where $m_a = \frac{v^2  a^2}{M_A}$ and $m_b = \frac{v^2  b^2}{M_B}$ can in
general be complex. We assume $|m_b| \ll |m_a|$,
which can originate from a hierarchy in $M_A$ and $M_B$, in the parameters
$a$ and $b$, or a combination of both.
This is nothing but sequential dominance.

The scale of family symmetry breaking,
the messenger scale $\Lambda$ and the scale of
the right-handed neutrino masses is undetermined
and can be chosen appropriately. This is due to
the fact, that
\begin{equation}
 M_\nu \sim v^2 \frac{\langle \phi \rangle^2}{M_R \Lambda^2} \;,
\end{equation}
from which this freedom of choice is obvious.

Clearly, the unitary matrix that diagonalises $M_\nu$
depends on only one complex parameter 
\be
  \frac{m_b}{m_a} ~=~\epsilon \,e^{i\alpha}\ , \qquad \epsilon,\alpha \in
  \mathbb R \ .\label{expan}
\ee 
An overall Majorana phase related to the parameter $m_a$ can be set to zero
without loss of generality. With $|m_b|\ll |m_a|$ we can use $\epsilon$ as an
expansion parameter for the resulting neutrino masses and mixing angles. We
have computed our results to second order in $\epsilon$. General
formulas can be found in \cite{King:1998jw}. 

Before determining the PMNS matrix explicitly, it is worth to show why the
neutrino mass matrix $M_\nu$ in Eq.\ \eqref{numassmatrix} implies the TM$_1$
mixing form in Eq.\ \eqref{TM1} where the first column is proportional to
$(-2,1,1)^T$. The reason is simply that $\lv\phi_3^{\nu}\rv\propto (-2,1,1)^T$
is an eigenvector of $M_\nu$ in Eq.\ \eqref{numassmatrix} with a zero
eigenvalue corresponding to the first neutrino mass being zero. The reason for
this is that $M_\nu$ in Eq.\ \eqref{numassmatrix} is a sum of two terms, the
first being proportional to $\lv\phi_1^{\nu}\rv\lv{\phi_1^{\nu}}\rv^T$  and
the second being proportional to $\lv\phi^{}_{120}\rv\lv\phi_{120}^T\rv$.  
Since $\lv\phi_3^{\nu}\rv\propto (-2,1,1)^T$ is orthogonal to both
$\lv\phi_1^{\nu}\rv$ and $\lv\phi_{120}\rv$ it is then clearly annihilated by
the neutrino mass matrix, i.e. it is an eigenvector with zero
eigenvalue. Therefore we immediately expect $M_\nu$ in
Eq.\ \eqref{numassmatrix} to be diagonalised by the TM$_1$ mixing matrix where
the first column is proportional to $\lv\phi_3^{\nu}\rv\propto
(-2,1,1)^T$. Although the remainder of this subsection gives a perturbative
diagonalisation of $M_\nu$ in Eq.\ \eqref{numassmatrix}, we already know it
must lead to TM$_1$ mixing exactly to all orders according to this general argument.

Mindful of the PDG phase conventions for the PMNS mixing matrix we write
\be
U_{\mathrm{PMNS}}^T  P' M_\nu P' U_{\mathrm{PMNS}}^{} ~=~
M_\nu^{\mathrm{diag}} ~=~
\mathrm{diag}(0,m^\nu_2,m^\nu_3) \ ,\label{meff}
\ee
with
\be
U_{\mathrm{PMNS}} ~=~ 
\begin{pmatrix}
1&0&0\\
0&c_{23}&s_{23}\\
0&-s_{23}&c_{23}
\end{pmatrix}
\begin{pmatrix}
c_{13}&0&s_{13}e^{-i\delta}\\
0&1&0\\
-s_{13}e^{i\delta}&0&c_{13}
\end{pmatrix}
\begin{pmatrix}
c_{12}&s_{12}&0\\
-s_{12}&c_{12}&0\\
0&0&1
\end{pmatrix}
 P \ ,
\ee
where $c_{ij}=\cos\theta_{ij}$ and $s_{ij}=\sin\theta_{ij}$. Furthermore,
$P=\mathrm{diag}(1, e^{i\frac{\alpha_{2}}{2}},e^{i\frac{\alpha_{3}}{2}})$ is
the Majorana phase matrix and  $P'=\mathrm{diag}(e^{i\delta_e},
e^{i\delta_\mu},e^{i\delta_\tau})$ is an unphysical phase matrix which is
required to bring the PMNS matrix into PDG form. 

In order to determine $U_{\mathrm{PMNS}}$ we first calculate the eigenvalues
of $(M_\nu^\dagger M_\nu)/m_a^2$ which will be functions of $\epsilon$ and
$\alpha$. Requiring orthogonality, the corresponding eigenvectors can be
obtained analytically. These eigenvectors, normalised to the unit length,
comprise the unitary matrix which diagonalises $M_\nu^\dagger M_\nu$ and thus
also $M_\nu$. It can therefore be identified with $P'U_{\mathrm{PMNS}}$ once
the Majorana phases have been adjusted to give real masses in
Eq.~\eqref{meff}. Using $\epsilon$ as our expansion parameter we obtain the
second order result,\footnote{We remark that the CP violating Dirac phase
  $\delta$ is only determined to order $\epsilon$ as it always appears
  together with $\sin\theta_{13}$ which in turn is already of order
  $\epsilon$. For completeness we also list the results for the unphysical
  phases $\delta_e=- \frac{\epsilon}{2}\sin\alpha (1-5\epsilon\cos\alpha)$, 
$\delta_\mu = - \frac{\epsilon}{2}\sin\alpha (3-7\epsilon\cos\alpha)$, and
$\delta_\tau = \pi + \frac{\epsilon}{2}\sin\alpha (1-\epsilon\cos\alpha)$.}
\bea
\label{analytic1}
m_2^\nu&=& \Big[ 3\,  \epsilon ~-~3\,  \epsilon^2 \cos \alpha \Big] m_a\ ,\\
m_3^\nu&=&  \Big[2~+~2 \, \epsilon \, \cos\alpha ~+~ \frac{\epsilon^2}{2} \,(7-\cos
2\alpha)\Big] m_a \ ,\\
\theta_{23} &=& \frac{\pi}{4} ~+~ \epsilon \cos \alpha ~+~ \epsilon^2
\left(\frac{3}{2}-\cos 2\alpha \right) \ ,\label{theta23} \\
\theta_{12} &=& \arcsin \frac{1}{\sqrt{3}} ~-~
\frac{\epsilon^2}{2\sqrt{2}}\ ,\\
\theta_{13} &=& \frac{\epsilon}{\sqrt{2}}  ~+~\frac{\epsilon^2}{2\sqrt{2}}\,
\cos\alpha \ , \\
\delta &=& \alpha ~-~ \epsilon\, \frac{5}{2}\, \sin\alpha ~~~
~~~ \text{ (only up to order } \epsilon)\ ,\\
\alpha_{2} &=& -\alpha~+~2\,\epsilon\, \sin\alpha ~-~3\,\epsilon^2 \sin 2\alpha \ ,\\
\label{analytic2}
\alpha_{3} &=& 0 \ .
\eea
Note that the PMNS matrix has only one non-trivial Majorana phase as one of
the neutrinos is exactly massless. These results are only slightly modified if
we choose the $(1,0,2)^T$ alignment for the subdominant neutrino term: 
$\theta_{23} \rightarrow \frac{\pi}{2}-\theta_{23}$, 
$\delta  \rightarrow \pi+ \delta $,
$\delta_e \rightarrow \pi+\delta_e$, and
$\delta_\mu \leftrightarrow \delta_\tau$.
All observables in the neutrino sector can be expressed in terms of $m_a,
\epsilon$ and $\alpha$. Excluding Majorana phases (and the mass of the
massless neutrino), this means that the model class makes three predictions
which should  be testable in future oscillation experiments since $\theta_{13}$
is comparatively large.  

It is useful to compare the above predictions to a general leading order
parametrisation of the PMNS mixing matrix in the PDG convention in terms of
deviations from TB mixing \cite{King:2007pr},  
\begin{eqnarray}
U_{\mathrm{PMNS}} =
\left( \begin{array}{ccc}
{\frac{2}{\sqrt{6}}}(1-\frac{1}{2}s)  & \frac{1}{\sqrt{3}}(1+s) & \frac{1}{\sqrt{2}}re^{-i\delta } \\
-\frac{1}{\sqrt{6}}(1+s-a + re^{i\delta })  & \phantom{-}\frac{1}{\sqrt{3}}(1-\frac{1}{2}s-a- \frac{1}{2}re^{i\delta })
& \frac{1}{\sqrt{2}}(1+a) \\
\phantom{-}\frac{1}{\sqrt{6}}(1+s+a- re^{i\delta })  & -\frac{1}{\sqrt{3}}(1-\frac{1}{2}s+a+ \frac{1}{2}re^{i\delta })
 & \frac{1}{\sqrt{2}}(1-a)
\end{array}
\right)P\ ,~~
\end{eqnarray}
where the deviation parameters $s,a,r$ are defined as \cite{King:2007pr}, 
\be
 \sin \theta_{12} = \frac{1}{\sqrt{3}}(1+s)\ , \qquad
\sin \theta_{23}  =  \frac{1}{\sqrt{2}}(1+a)\  , \qquad 
 \sin \theta_{13} =   \frac{r}{\sqrt{2}}\ .
\label{rsa}
\ee
At leading order the above predictions can be expressed by
\be
a\,=\,r \cos\delta \ , \qquad s\,=\,0 \ ,
\ee
where
\be
r \,=\, \frac{2}{3} \, \frac{m^\nu_2}{m^\nu_3}  
\,\sim \,  \frac{2}{15} ~~ \rightarrow ~~ 
\theta_{13} \sim~5^\circ-6^\circ \ ,
\ee
where the predicted reactor angle may be compared to
Eq.\ \eqref{reactor}.\footnote{Note that in a model where the charged lepton
  mass matrix is not diagonal, one must combine the charged lepton corrections
  with the underlying TB neutrino mixing deviations to formulate the total
  observed deviation from TB mixing as discussed in \cite{Antusch:2008yc}.} 
We emphasise that these predictions hold true for both the $(1,2,0)^T$ as well
as the $(1,0,2)^T$ alignment. In both cases, with a suitable choice of phase
convention, the leading order mixing matrix can be written in the form,  
\begin{eqnarray}
U_{\mathrm{TM_1}} =
P' \left( \begin{array}{ccc}
{\frac{2}{\sqrt{6}}}  & \frac{1}{\sqrt{3}} & \frac{1}{\sqrt{2}}re^{-i\delta } \\
-\frac{1}{\sqrt{6}}  & \frac{1}{\sqrt{3}}(1- \frac{3}{2}re^{i\delta })
&\phantom{-} \frac{1}{\sqrt{2}}(1+re^{-i\delta }) \\
-\frac{1}{\sqrt{6}}& \frac{1}{\sqrt{3}}(1+ \frac{3}{2}re^{i\delta })
 & -\frac{1}{\sqrt{2}}(1-re^{-i\delta })
\end{array}
\right)P\ ,~~
\label{PMNS1}
\end{eqnarray}
where Eq.\ \eqref{PMNS1} corresponds to a small angle expansion of TM$_1$
mixing in Eq.\ \eqref{TM1}. However, from the general argument given earlier
in this subsection, we expect TM$_1$ mixing in Eq.\ \eqref{TM1} to be valid to
all orders beyond the small angle approximation.

\subsection{Numerical results}

\begin{figure}
\centering
\includegraphics[scale=0.55]{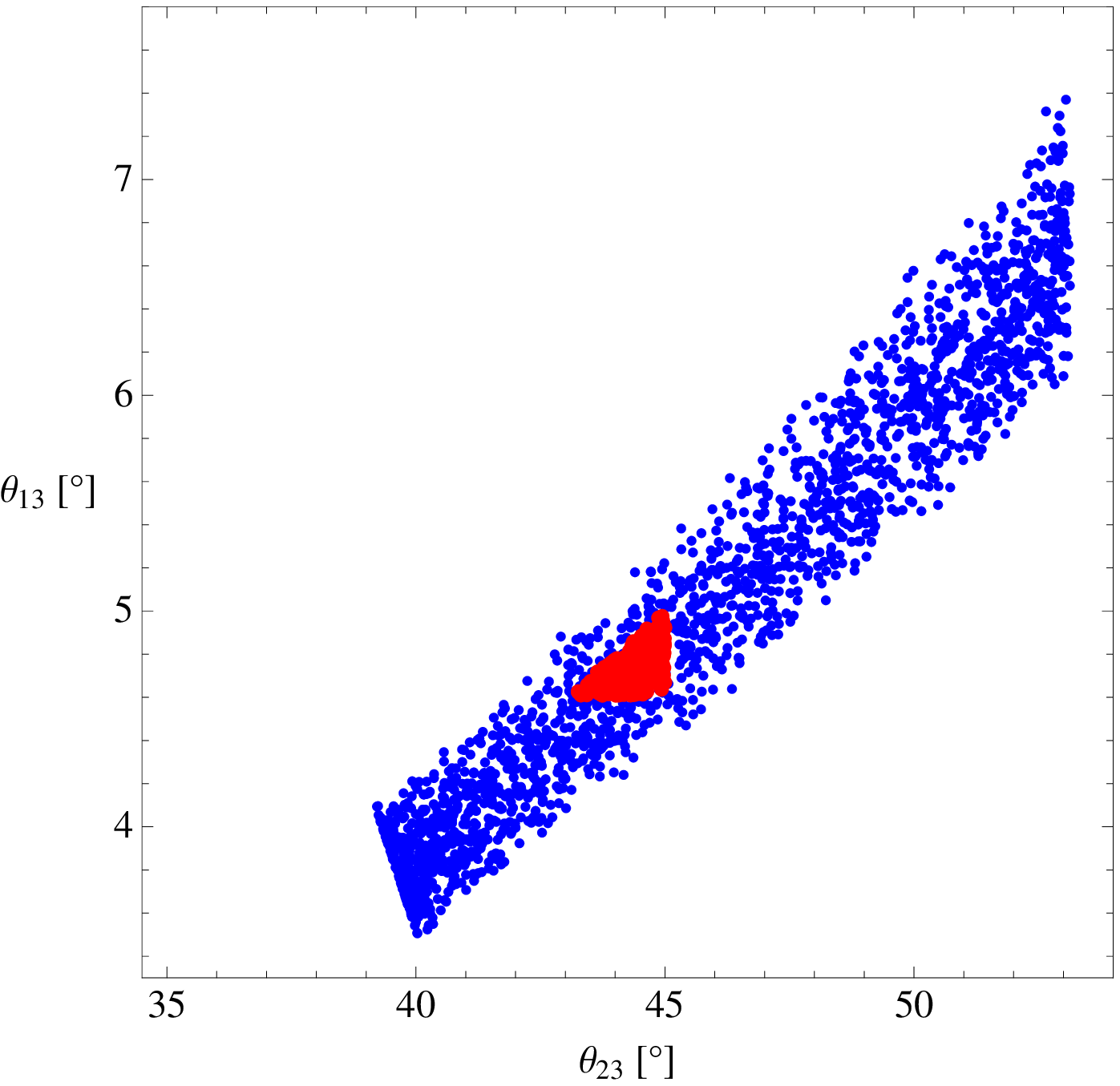}
\includegraphics[scale=0.55]{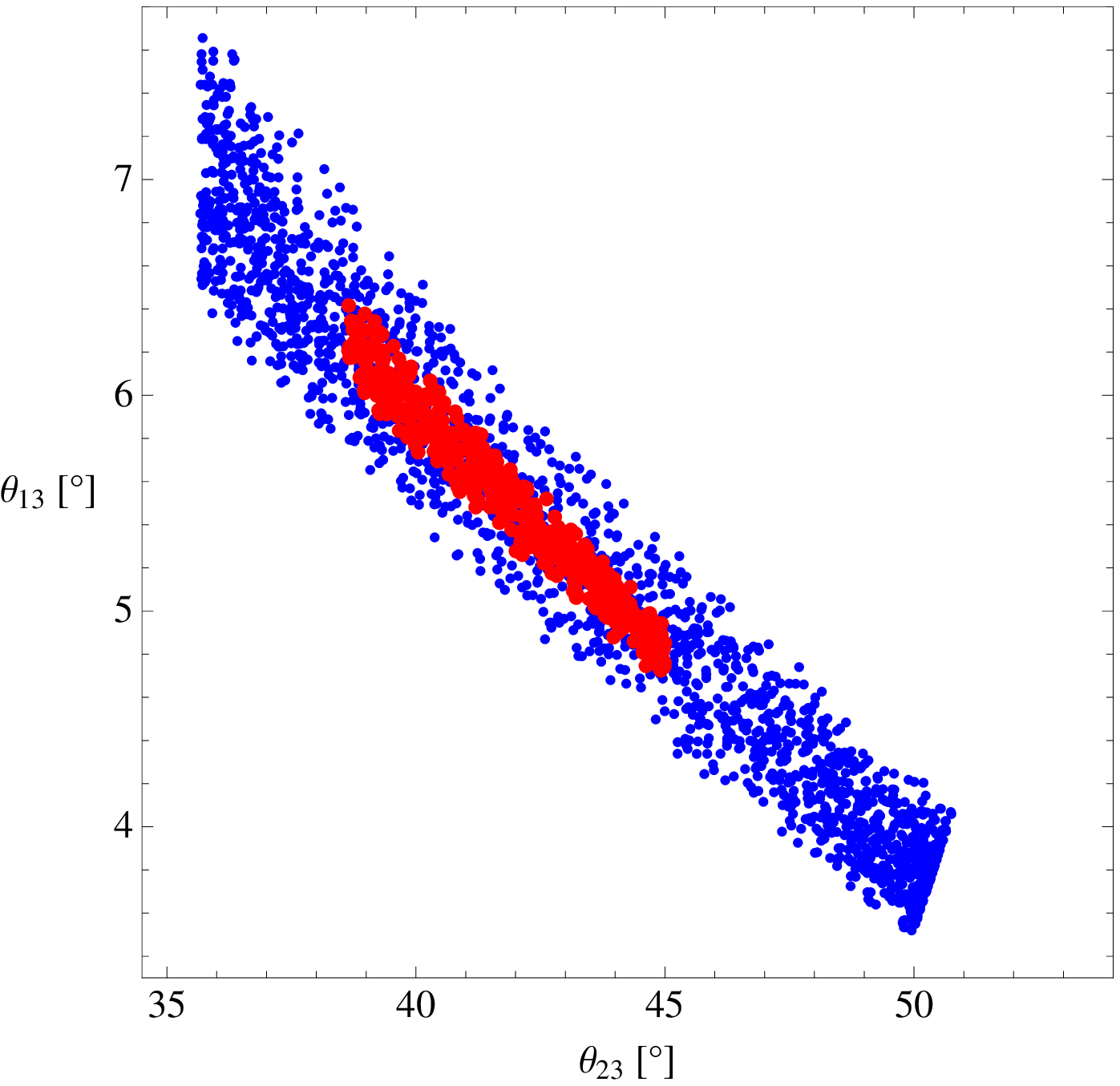}
\includegraphics[scale=0.55]{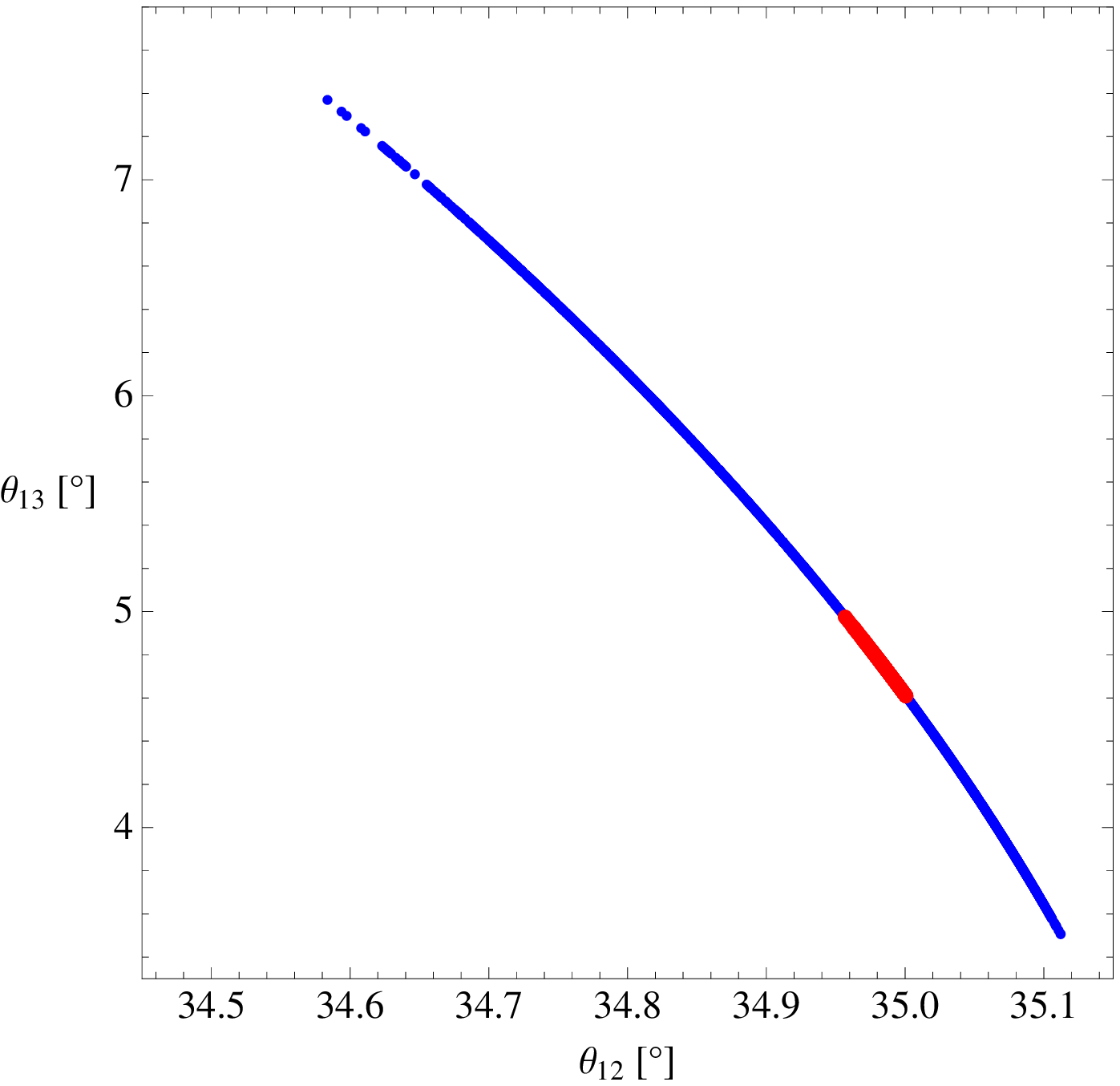}
\includegraphics[scale=0.55]{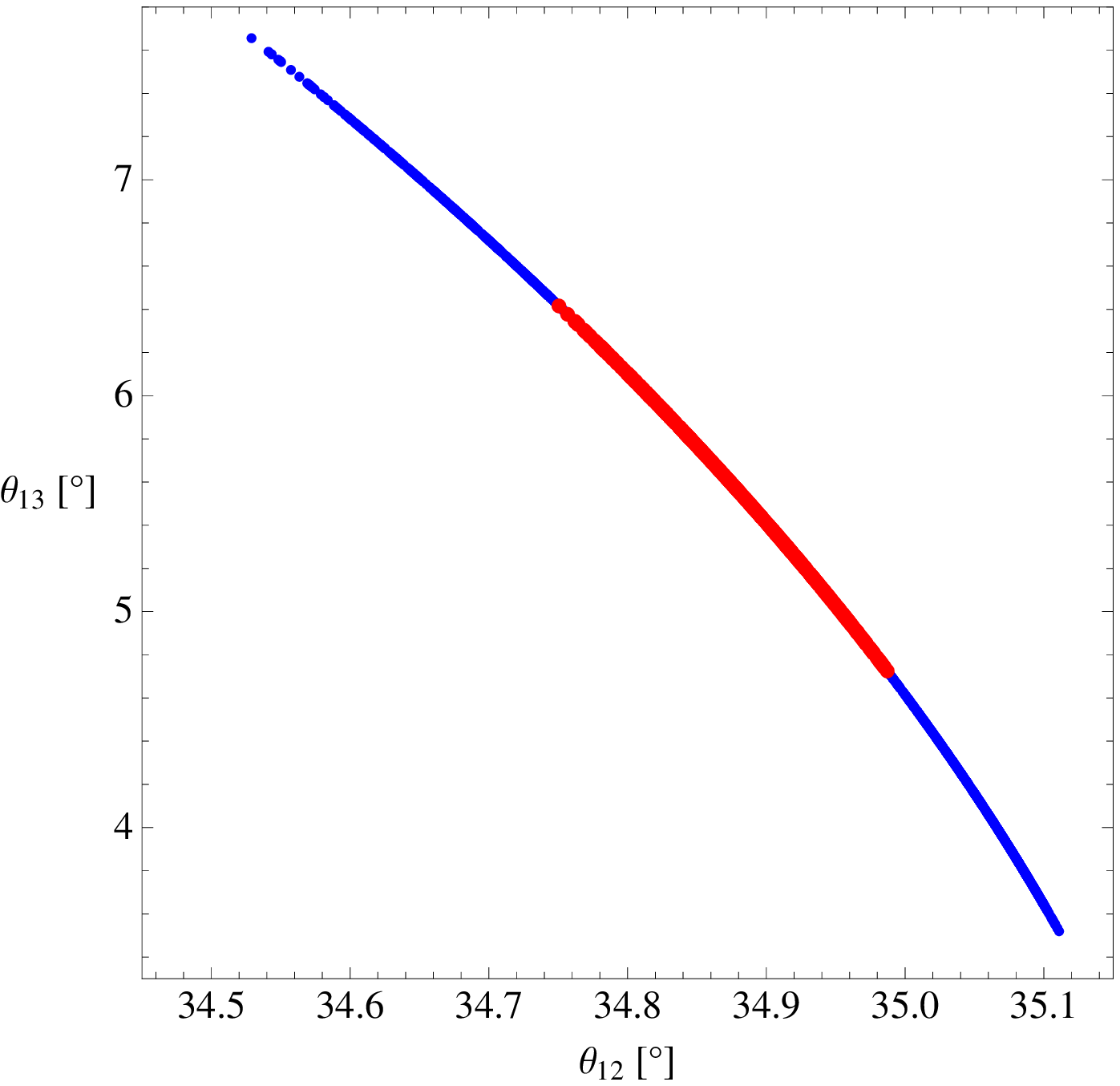}
\caption{The correlations between $\theta_{13}$ and the other two mixing angles
  in CSD2. The panels on the left/right show the results for the
  $(1,2,0)^T$/$(1,0,2)^T$ alignment. The regions compatible with the 1$\sigma$
  (3$\sigma$) ranges of the atmospheric and solar neutrino mass squared
  differences and mixing angles, taken from \cite{Fogli:2011qn}, are depicted
  by the red (blue) points.\label{Fig:Correlations1}}
\end{figure}

\begin{figure}
\centering
\includegraphics[scale=0.55]{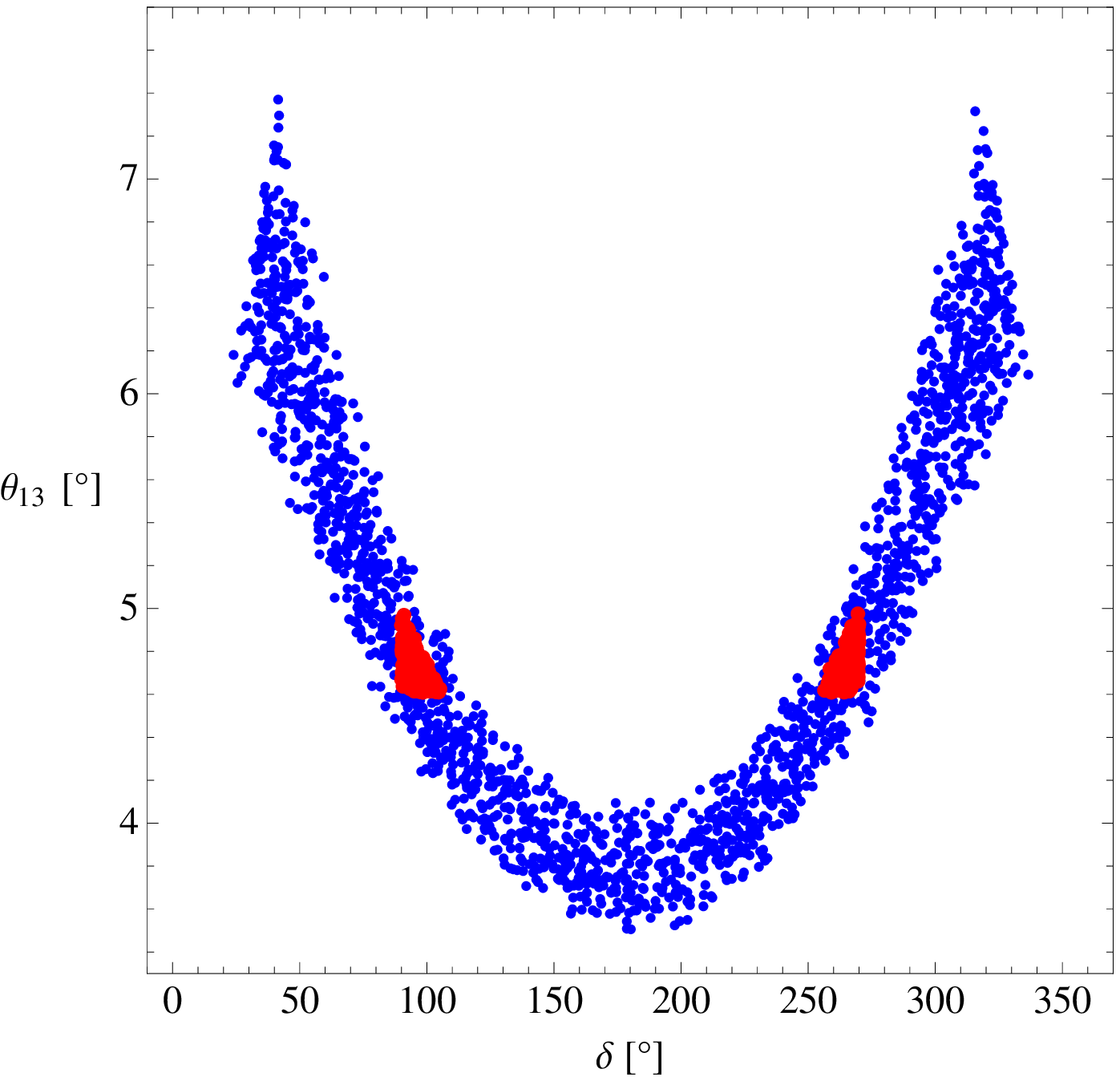}
\includegraphics[scale=0.55]{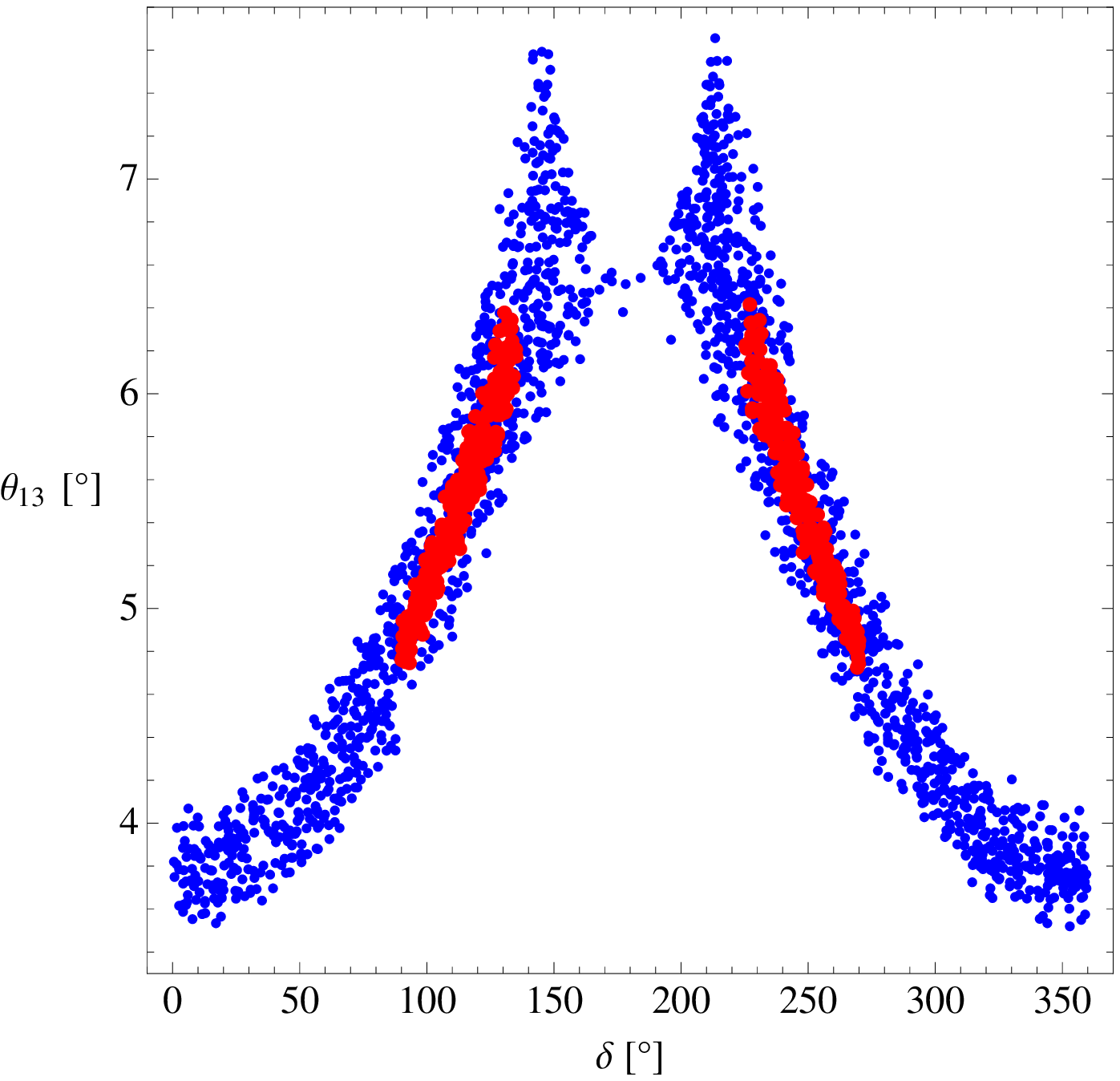}
\includegraphics[scale=0.55]{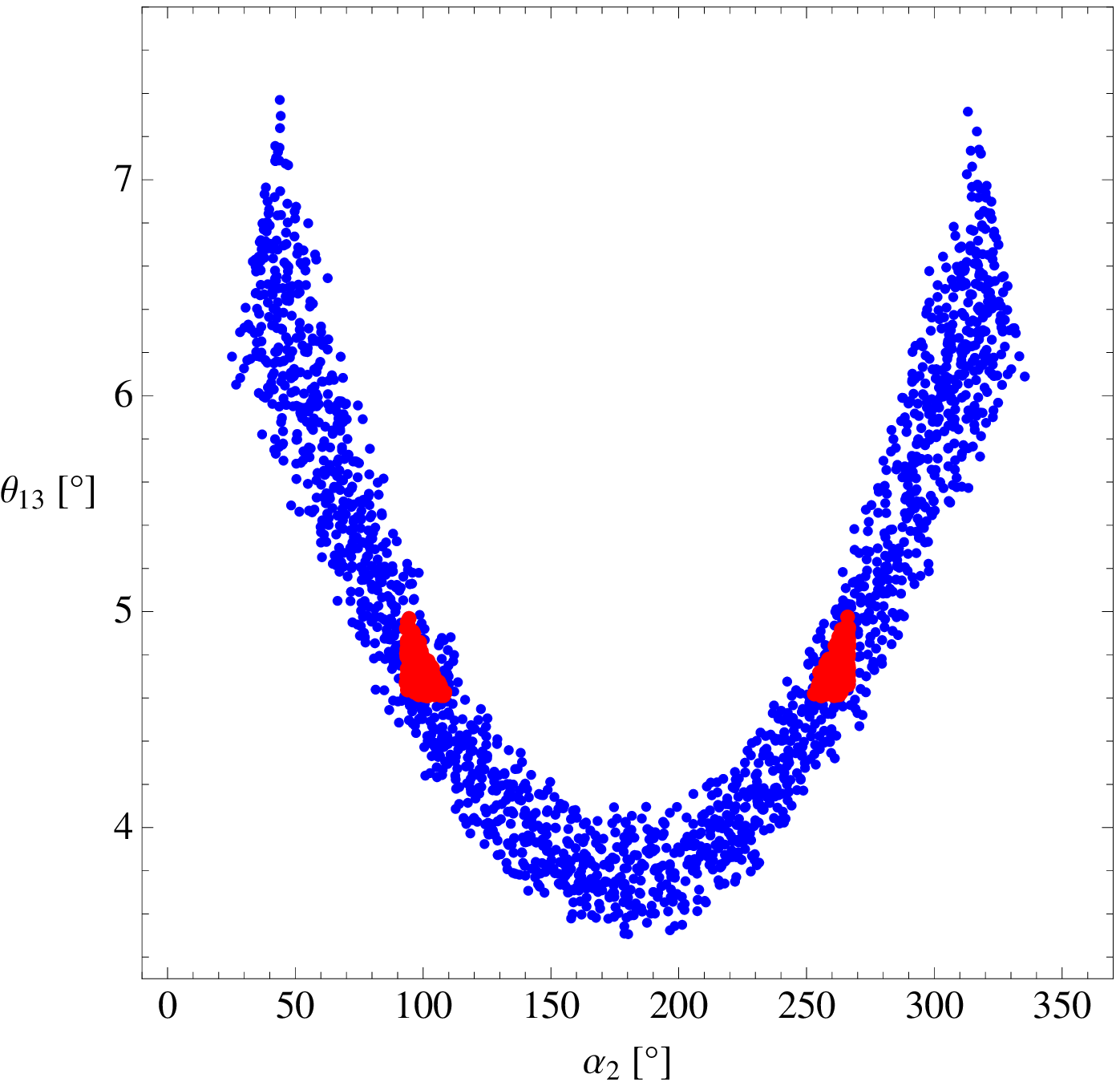}
\includegraphics[scale=0.55]{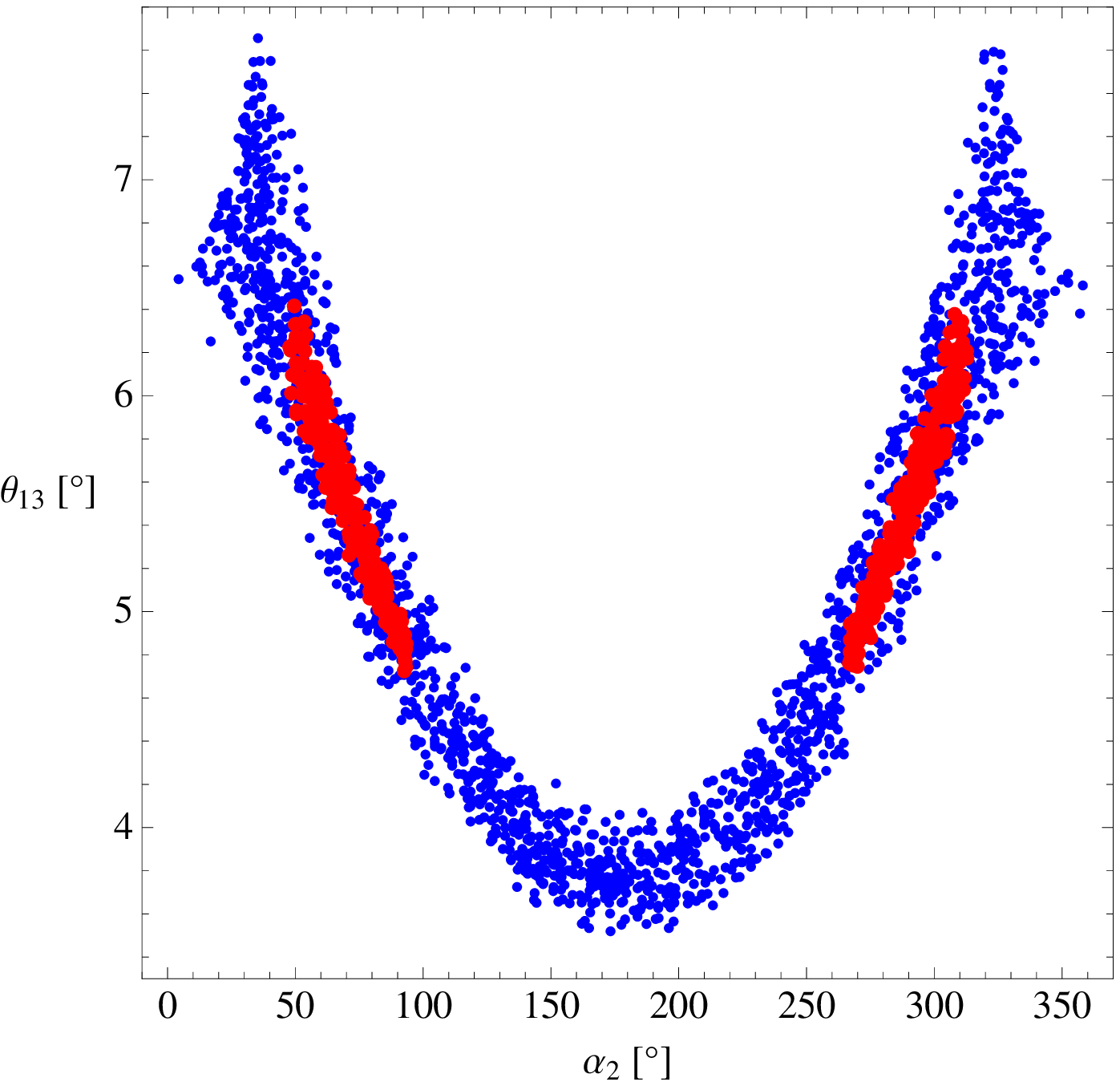}
\caption{The correlations between $\theta_{13}$ and the two physical phases in
  CSD2. The panels on the left/right show the results for the
  $(1,2,0)^T$/$(1,0,2)^T$ alignment. The regions compatible with the 1$\sigma$
  (3$\sigma$) ranges of the atmospheric and solar neutrino mass squared
  differences and mixing angles, taken from \cite{Fogli:2011qn}, are depicted
  by the red (blue) points.\label{Fig:Correlations2}}
\end{figure}

In this section we present the numerical results for the two CSD2 cases
defined by Eqs.\ \eqref{eq:NeutrinoYukawas} and \eqref{eq:MR} in
Figs.\ \ref{Fig:Correlations1} and \ref{Fig:Correlations2}.
We used random values for $m_a$, $m_b$ and $\alpha$ as input
and calculated with the Mixing Parameter Tools provided with the
REAP package \cite{Antusch:2005gp} the resulting neutrino masses,
mixing angles and CP violating phases.  We used then the recent global
fit results from \cite{Fogli:2011qn} for the solar and atmospheric neutrino mass
squared differences and mixing angles deduced from the new reactor
fluxes as a constraint. We have also checked that the numerical results
agree well with the second
order analytic results in Eqs.~(\ref{analytic1}-\ref{analytic2}).

There are some interesting features of the plots. First of all, note that
$\theta_{13}$ can go up to more than 7$^\circ$ in the 3$\sigma$ interval. It
is, however, more interesting to look at the 1$\sigma$ regions. The
atmospheric mixing angle $\theta_{23}$ has an upper 1$\sigma$ bound of
45$^\circ$, which is very restrictive for the $(1,2,0)^T$ alignment. Indeed, by
this bound, $\theta_{13} > 5^\circ$ is disfavoured in the $(1,2,0)^T$ case, while
for the $(1,0,2)^T$ alignment values up to $6.4^\circ$ are still allowed, see
upper panels of Fig.\ \ref{Fig:Correlations1}. This is due to the fact that the deviations
from $\theta_{23} = 45^\circ$ have opposite signs for both cases.  Turning to
the solar mixing angle, the 1$\sigma$ region for $\theta_{12}$ induces a lower
bound on $\theta_{13}$ of approximately $4.5^\circ$, which is identical in
both cases, see lower panels of Fig.\ \ref{Fig:Correlations1}. 

It is also interesting to look at the phases in Fig.\ \ref{Fig:Correlations2}. 
In the $(1,2,0)^T$ alignment case a phase difference $\alpha$,
cf.~Eq.~\eqref{expan}, of approximately $90^\circ-100^\circ$ or $260^\circ -
270^\circ$ is preferred as can be seen in the upper left panel of
Fig.\ \ref{Fig:Correlations2}.\footnote{Keep in mind that the Dirac CP phase
  is almost identical to the phase difference $\alpha$ in this case.}  For the
$(1,0,2)^T$ alignment, the preferred values of the Dirac CP phase span bigger
regions, but still the CP conserving case is not preferred, see upper right
panel of Fig.\ \ref{Fig:Correlations2}. Actually, the maximally CP violating
cases $\delta = \pm 90^\circ$ are in both cases at the edge of the
preferred regions. This is due to the fact that the corrections to a
maximal atmospheric mixing are very small for $\alpha\approx
\delta=\pm90^\circ$, see Eq.~\eqref{theta23}. Such a phase can emerge naturally in models with spontaneous CP violation from discrete symmetries \cite{Antusch:2011sx}.

\section{\label{4}PMNS-leptogenesis link}
As has been noticed in
\cite{Antusch:2006cw,Choubey:2010vs,AristizabalSierra:2009ex}, in models where
TB mixing is realised via flavons which are orthogonal to each other, as in
CSD or more generally in scenarios which satisfy the conditions of form
dominance \cite{Chen:2009um,Choubey:2010vs}, the CP asymmetries for
leptogenesis vanish.

On the contrary, in models with $(1,2,0)^T$ or $(1,0,2)^T$ vacuum alignment -
since the flavon vevs of the model are now no longer orthogonal - the
asymmetry does not vanish, rendering models of this type attractive for
cosmology. 

Furthermore, the two zero textures in $Y_\nu$ imply a direct link
between the CP violation for leptogenesis and the Dirac CP phase $\delta$, as
has been discussed for models with sequential dominance and hierarchical RH neutrinos in
\cite{King:2002qh, Ibarra:2003up, Antusch:2006cw}. The produced baryon asymmetry $Y_B$ from
leptogenesis in models with $(1,2,0)^T$ or $(1,0,2)^T$ vacuum alignment satisfies 
\bea\label{eq:BAU}
Y_B \propto \pm\sin \delta \: ,
\eea
meaning that a measurement of $\delta$ in future neutrino oscillation
experiments allows to draw conclusions about the prospects for leptogenesis.  

The sign in Eq.~(\ref{eq:BAU}) depends on the
choice of the new flavon alignment, either $(1,2,0)^T$ or $(1,0,2)^T$, as well as
on which of the RH neutrinos is the lightest, $N_1$ 
with mass $M_A$ or $N_2$ with mass $M_B$ (cf.~\cite{Antusch:2006cw}).
Explicitly, the ``$+$'' sign applies to the $(1,0,2)^T$ alignment with $M_A \ll M_B$
and to the $(1,2,0)^T$ alignment with $M_B \ll M_A$. The ``$-$'' sign holds for the
other two cases, the $(1,0,2)^T$
alignment with $M_B \ll M_A$ and the $(1,2,0)^T$ alignment with $M_A \ll M_B$.

Since the baryon asymmetry $Y_B$ is positive, it follows that, in models with a fixed
alignment and RH neutrino masses, leptogenesis requires $\delta$ in a specific range.  
In models where the ``$+$'' sign applies in Eq.~(\ref{eq:BAU}) only the region around 
$\delta = 90^\circ$ generates the correct positive $Y_B$, while in models where the 
``$-$'' sign holds only the region around $\delta = 270^\circ$ is valid.

Current global fits \cite{Schwetz:2011zk} favour $\sin \delta$ being negative and this suggests that the negative sign is favoured in Eq.\ \eqref{eq:BAU}. According to the above discussion this suggests either 
the $(1, 0, 2)^T$ alignment with $M_B \ll  M_A$ or the $(1, 2, 0)^T$ alignment with $M_A \ll  M_B$. The latter possibility corresponds to so called ``light sequential dominance'' which plays a special role in leptogenesis
within the framework of two right-handed neutrino models as recently
discussed in \cite{Antusch:2011nz}.

\section{\label{5}Summary and conclusions}

Recently T2K have published evidence for a large non-zero reactor angle which,
if confirmed, would exclude the tri-bimaximal mixing pattern. In this paper we
have presented a model which fixes the reactor angle while preserving
trimaximal solar mixing.  In particular we have shown how a variant of
trimaximal mixing, called TM$_1$ mixing in Eq.\ \eqref{TM1} with the solar
angle given by $\sin\theta_{12} \approx 1/\sqrt{3}$, results from an extension
of constrained sequential dominance involving  new vacuum alignments along
the $(1,2,0)^T$ or $(1,0,2)^T$ directions in flavour space. We have shown that
such alignments are naturally achieved using orthogonality, and may  replace
the role of the subdominant flavon alignment $(1,1,1)^T$ in constrained
sequential dominance.  We have proposed the first model in the literature of
this kind  leading to TM$_1$ mixing where the reactor angle is related to the
ratio of the solar to the atmospheric neutrino masses, $\theta_{13} =
\frac{\sqrt{2}}{3} \, \frac{m^\nu_2}{m^\nu_3}$.
We emphasise that the considered model is merely representative of a
general strategy based on CSD2 for obtaining TM$_1$ mixing together with the
above prediction for the reactor angle.

We have studied the phenomenological consequences of CSD2 both analytically
and numerically. The analytic treatment confirms that TM$_1$ mixing results,
at leading order, in a reactor angle which is predicted to be proportional to
the ratio of the solar to the atmospheric neutrino masses, yielding
$\theta_{13} \sim 5^\circ - 6^\circ$, while the atmospheric angle is given by 
the sum rule $\theta_{23} \approx 45^\circ + \sqrt{2} \theta_{13} \cos\delta$, 
where the leptonic Dirac CP phase $\delta$ is undetermined by CSD2,
but experimentally preferred to lie in a range of $90^\circ-130^\circ$ or $230^\circ-270^\circ$.
The numerical results agree well with the second order analytic results,
and demonstrate the full range of neutrino mixing parameters possible with CSD2,
although this range could be extended in models which contain additional contributions from
charged lepton mixing.

Finally we have seen that in CSD2 leptogenesis is unsuppressed due to the
violation of form dominance and that the decay asymmetries feature a
direct link between the CP phase for leptogenesis and the Dirac CP phase $\delta$, with the produced baryon asymmetry $Y_B \propto \pm\sin \delta$.

In conclusion, CSD2 leads to a highly predictive form of leptonic mixing,
with the solar angle
tightly constrained to its trimaximal value, the reactor angle predicted
to be within the
range of recent global fits, and the atmospheric angle correlated with the
Dirac CP phase $\delta$ which
is precisely equal to the leptogenesis phase.
The large reactor angle indicated  by T2K therefore opens up the exciting
possibility of an early measurement of
the low energy CP violating phase $\delta$ which is also responsible for
the matter-antimatter asymmetry
of the Universe within this class of models.

\section*{Acknowledgments}
S.~A.\ acknowledges partial support by the DFG cluster of excellence
``Origin and Structure of the Universe.''
S.~F.~K.\ and C.~L.\ acknowledge support from the STFC Rolling Grant
No. ST/G000557/1 and the EU ITN grant UNILHC 237920 (Unification in the LHC era).
M.~S.\ acknowledges partial support from the Italian government under the project number PRIN 2008XM9HLM ``Fundamental interactions in view of the Large Hadron Collider and of astro-particle physics''.

\end{document}